\documentclass[aps,prb,final,twocolumn,showpacs]{revtex4}

\usepackage{graphicx}
\usepackage{dcolumn}
\usepackage{bm}
\usepackage{color}

\begin{document}

\title{Dynamical gap generation in graphene nanoribbons: An effective relativistic field theoretical model}

\author{A. J. Chaves$^1$, G. D. Lima$^2$, W. de Paula$^1$,  C. E. Cordeiro$^{3}$, A. Delfino$^{3}$, T. Frederico$^{1}$,  and O. Oliveira$^{1,4}$}

\affiliation{
{\it $^{1}$Departamento de F\'\i sica, Instituto Tecnol\'ogico de Aeron\'autica,
12228-900 S\~ao Jos\'e dos Campos, Brazil \\
$^2$Departamento de F\'isica, Universidade Federal do Piau\'i, 64049-550
Teresina, Piau\'i, Brasil \\
$^{3}$Instituto de F\'\i sica, Universidade Federal Fluminense, Avenida
Litor\^anea s/n, 24210-340 Niter\'oi, RJ, Brazil \\
$^{4}$Departamento de F\'isica, Universidade de Coimbra, 3004 516 Coimbra,
Portugal}}

\date{\today}

\begin{abstract}
We show that the assumption of a nontrivial zero band gap for a graphene sheet within an effective relativistic field theoretical model description of interacting Dirac electrons on the surface of graphene describes the experimental band gap of graphene nanoribbons for a wide range of widths. The graphene band gap is dynamically generated, corresponding to a nontrivial gapless solution, found in the limit of an infinitely wide graphene ribbon. The nanoribbon band gap is determined by the experimental graphene work function.
\end{abstract}


\maketitle

\section{Introduction}
Graphene sheets, carbon nanotubes (CNTs)
and graphene nanoribbons (GNRs) are very closely related
subjects governed by quantum mechanical effects in a constrained
two-dimensional (2D) surface. In this compactified
dimensional world, symmetry and scale drive the astonishing
properties that these one-atom-thick materials exhibit. GNRs
are recognized today as promising nanosheets to be used
in nanoelectronic and spintronic devices\cite{geim}. Such material has attracted a lot of attention since free-standing graphene sheets
were reported by using different experimental techniques. Experiments confirmed that the charge carriers can be described
as massless Dirac fermions\cite{massless}, opening a solid basis for its
relativistic treatment. It is expected that massless fermions
moving through graphene have an approximate ballistic transport
behavior with very small resistance due backscattering
suppression\cite{abanin}. Graphene is also a good thermal conductor,
when compared with other semiconductor junctions, due to
the high mobility of carriers at room temperature\cite{saitobook,areshkin}.

Energy band gaps for GNRs weremeasured in Refs. \cite{xiaolin} and \cite{melinda}
using different methods. First principles {\it ab initio} methods
\cite{ywson}, density functional theory (DFT)\cite{dft-vbarone}, and tight
binding (TB) models \cite{reich} were applied to calculate GNR band
gaps for a reasonable range of not too small widths. In opposition
to TB models, first-principles calculations do not support metallic
nanoribbons \cite{ywson}. On the other hand DFT calculations predict
energy gap oscillations as a function of the GNR for very small
widths \cite{dft-vbarone}.

The two-dimensional (2D) graphene sheet is essentially a zerogap
semiconductor (with the Fermi level $E_F$ precisely at $E$ = 0),
with a linear chiral carrier kinetic energy dispersion relation
given by $E=\hbar \, v_F \, k$ \cite{massless}, where k is the 2D
carrier wave vector and $v_F$ is the Fermi velocity (independent of
carrier density).

The fermion behavior on the graphene sheet has been discussed within
the free hamiltonian, $\hat{H_{o}}$, picture using four different
approaches \cite{geim2}:
(i) as Schr\"odinger fermions $\hat{H_{o}}=
\hat{p}^{2}/2m^{*}$, where $m^{*}$ is an effective mass; (ii) as
ultra-relativistic Dirac fermions, where $\hat{H_{o}}= c \vec{\sigma}\hat{ p}$; (iii)
as massless Dirac fermions with  $\hat{H_{o}}= v_{F} \, \vec{\sigma} \, \hat{p}$, where
$\vec{\sigma}$ are the Pauli matrices; and (iv) as massive chiral fermions
with  $\hat{H_{o}}= \vec{\sigma} \, \hat{p}^{2}/2m^{*}$, where
$\vec{\sigma}$ is a pseudospin matrix describing the two sublattices of
the honeycomb lattices \cite{castroneto}.

This work is based on a relativistic effective field theory of
electrons containing a local four-fermion interaction. The effective
interaction Lagrangian parameterizes the actual forces acting on the
electrons.  It resumes the relevant dynamics from the atom-electron
and electron-electron interaction within the many-body system
(including e.g. electron-phonon interactions). The rationale to use
effective contact interactions  goes back to the claim
\cite{brack,furn} that the Hartree-Fock approximation of the single
particle energy of the many-fermion system with the point-coupling
interaction can be compared to the first order contribution of the
surface density in a density functional theory \cite{nobel}.

The relativistic
picture of interacting fermions has to be understood
together with the concept of an effective mass.
In the case of the graphene the effective mass vanishes,
corresponding to a gapless material.  A nonvanishing
effective mass for GNRs means that a gap is open.
Our model provides a mechanism for dynamical mass generation
for the nanoribbons, even for a gapless graphene.

We assume an homogeneous 2D system of electrons (or holes),
\cite{massless} where the free part of the electron Hamiltonian
corresponds to case (iii), i.e., $\hat{H_{o}}= v_{F}\vec{
\sigma} \hat{p}$. The two sublattices of graphene, associated to the
Dirac points K and K', are taken into account by introducing a
factor 2 in the electron multiplicity.

The model is based on a relativistic Lagrangian field theoretical
approach in which a four-fermion scalar and a vector point-like
interactions are considered and the many-fermion problem solved in
the Hartree-Fock approximation on a 2D surface. The model has only
two free parameters, one related to the strengths of the
four-fermion interaction and the Fermi momentum cut-off, $\Lambda$.
Exploring the connection of the model with graphene, i.e., by
assuming that graphene is a nontrivial gapless solution of the
dynamical equations, we are able to relate the strengths of the
interactions and cut-off with the graphene work function. This
procedure allows to  fix the parameters and calculate the gaps of
the nanoribbons without any further assumptions.

Our work is organized as follows. In section II, we present the
model and  derive the gap equations. In section III, we present the
solution of the gap equations for the nanoribbon and compare our
results with the experimental data. \cite{xiaolin} Our concluding
remarks are given in the last section.

\section{The Model}

The electron-hole dynamics is described by a relativistic quantum
field effective Lagrangian in a 2+1 space-time, where the
interaction has a pairwise and contact form, which generalizes a
nonrelativistic 2D model \cite{cordeiro},
\begin{equation}
\mathcal{L}=\bar{\Psi}(i
{\gamma}^{\mu}\partial_{\mu}-m)\Psi+G_s(\bar{\Psi}\Psi)^2+G_v(\bar{\Psi}{\gamma}^{\mu}\Psi)^2,
\label{lagrangiana}
\end{equation}
where $s$ and $v$ refer to scalar and vector couplings, respectively, and
$\Psi$ refers to fermion fields. In the following, we use a system of units
such that $\hbar=v_F=1$.

Our method of solving the model for the nanoribbon starts
from a nanotube geometry where the fermions are constrained to move on
the cylindrical surface. The appropriate coordinates are the position
along the nanotube symmetry axis $(z)$ and the angle in the transverse
plane $(\theta)$ - see Ref. {\cite{cordeiro} for details. The GNRs
are obtained by {\it unfolding } an infinite length carbon nanotube.
The resulting sheet is a GNR with a width  $W$, after eliminating
dangling bonds. The procedure to obtain GNRs from CNTs is described in
Ref. \cite{dft-vbarone}.

\begin{figure}[t]
\vspace{-0.3cm}
\includegraphics[scale=0.3]{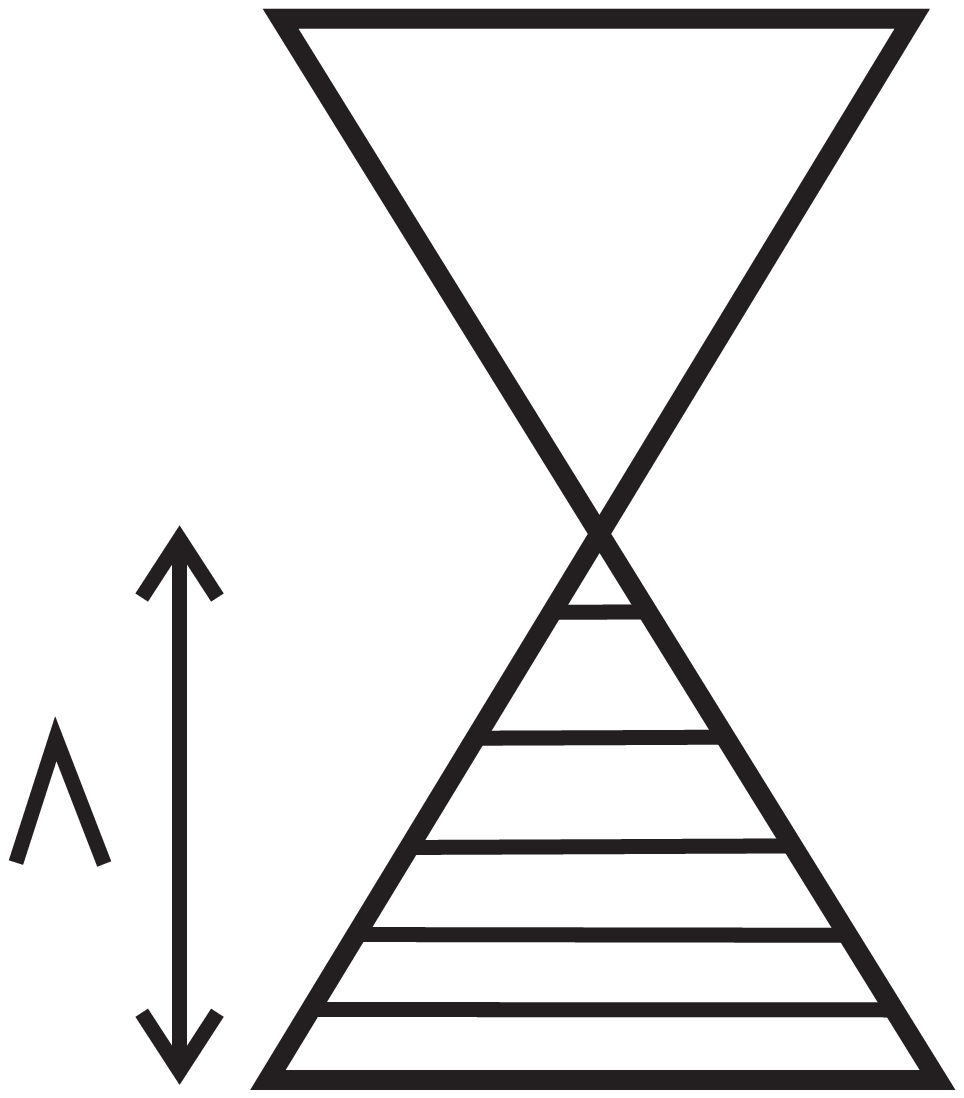}
\vspace{-0.3cm}
\caption{Schematic graphene bands for conducting (upper)
and valence (lower) fermions. For the discussion of $\Lambda$ see the text.  }\label{banda}
\end{figure}

The Dirac modes are quantized along the transversal direction after demanding
that the spinors vanishes at the edges of the nanoribbon. From this condition it
follows that the allowed fermion transverse momenta are given by $k_n = n \, \pi / W$, where
$n = \pm 1, \, \pm 2, \cdots , \, n_{max} $ with $n_{max} $ being the largest integer
smaller than $\Lambda  \, W / \pi$. The maximum momentum of the negative energy electrons
filling the valence band, as illustrated in Fig. \ref{banda}, is $\Lambda$.

The fermion mass, i.e. the electron self-energy, is
generated dynamically and is computed from the gap equations for the GNRs,
using a self-consistent Schwinger-Dyson equation at one-loop level (Hartree-Fock approximation).
Recall that the Schwinger-Dyson equation has a cutoff in the momentum of the negative electron energy in the
valence band given by $\Lambda$.
The gap equations for the self-energies, after integration on the longitudinal momentum, are given by
\begin{eqnarray}
       \Sigma_{s} & =  & \frac{2 \, G}{\pi W} \, (\Sigma_{s}+m)  \, \sum^{+ n_{max}}_ {n = - n_{max}} S(k_n) \, ,
       \label{sigmas} \\
     \Sigma_{v}^{0}  & = &  \frac{2 \, G \, \Lambda^2}{\pi^{2}} \, f(x) \, ,
     \label{sigmav0} \\
        \Sigma_v^{\theta} & = & -\frac{2 \, G}{\pi W} \,\sum^{+ n_{max}}_ {n = - n_{max}}
        S(k_n) \, (k_n-\Sigma_v^{\theta}) \, ,
       \label{sigmavtheta}
\end{eqnarray}
where
\begin{equation}
      S(k) = \ln\left[\sqrt{\frac{\Lambda^{2}-k^{2}}{a^2}}
      + \sqrt{\frac{\Lambda^{2}-k^{2}}{a^2}+1}\right] \, ,
\end{equation}
\begin{equation}
f(x) = \frac{1}{x}\sum^{+ n_{max}}_ {n = - n_{max}} \sqrt{1-\frac{n^2}{x^2}}
\label{fx}
\end{equation}
and $a^{2}=(k_{n}-\Sigma_v^{\theta})^2 + (\Sigma_{s}+m)^2$.

In Eqs. (\ref{sigmas}) to (\ref{sigmavtheta}) we define $x=\Lambda W/\pi$,  $G=G_{s} - 3 \, G_{v}$
and a factor of 2 was included to take into account for electrons at the K and K' points.
Note that the scalar $G_S$ and the vector $G_v$ coupling constants combine
into a single coupling constant $G$. Furthermore, the self-energies given by
Eqs. (\ref{sigmas})-(\ref{sigmavtheta}) include a rest fermion mass $m$. However,
hereafter we consider only the $m=0$ case.

The scalar self-energy  $\Sigma_s$ is identified with the half bandgap nannoribbon.
The self-energy vector part is actually a quadri-vector containing  space-time components.
However, due to symmetry only the time component given by  $\Sigma_v^{o}$ survives.
This component is directly related to the number of particles N through
\begin{equation}
N=\frac{2LW}{G}\Sigma_v^0,
\label{number}
\end{equation}
where $L$ is the longitudinal length of the GNR.

In a relativistic picture, the infinite graphene sheet ($W\rightarrow\infty$)
has no electrons in the conducting band. The valence band is fully occupied and there is no gap
between both bands, i.e. $\,\Sigma_{s}\,=\,0$. Then, it remains to parametrize only Eq. (\ref{sigmav0}), since
$\Sigma_s=0$ is a trivial solution for Eq. (\ref{sigmas}). Recall that we are
considering massless fermions $m=0$. Anyway, insofar considering very large but finite values for $W$,
(\ref{sigmas}) admits other solutions besides the trivial $\Sigma_s = 0$ and one has an opening
gap generated dynamically.

Let us look for the nontrivial solution of (\ref{sigmas}). This defines a unique coupling
constant, a function of $\Lambda$, $G = G_{crit}$, which will be called the critical graphene limit.
If one takes the continuum limit of Eq. (\ref{sigmas}) with $m = 0$,
after setting $y= k_{n}/\Lambda$ it follows that
\begin{equation}
\frac{2\Lambda G_{crit}}{\pi^2}\int_{0}^1 dy\,\,\,
ln\left(\sqrt{\frac{1}{y^2}-1}+\frac{1}{y}\right)=1.
\end{equation}
Given that the above integral is equal to $\pi/2$, then the critical graphene limit requires a
\begin{equation}
G_{crit}=\frac{\pi}{\Lambda}  \, .
\label{gcrit}
\end{equation}
The $\Lambda$ dependence can be eliminated in favor of the graphene
experimental work function value $W \!\! F$= 4.8 eV  \cite{suzuki}.
The work function is related with the chemical potential $\mu$,
which needs to be computed in a consistent thermodynamic way
\cite{cordeiro,march}. The pressure $P$, the density energy $\cal E$
of the system determine the chemical potential by
$\,\mu\,=(\,P\,+\,\cal E\,)/\sigma$, with $\sigma=N/A$ and where $A$
is the area of the GNR. $P$ and $\cal E$ are given by the diagonal
terms of the energy-momentum tensor constructed from the model
Lagrangian (\ref{lagrangiana}) in the usual way. In a mean field
approach \cite{walecka},
\begin{eqnarray}
    {\cal E} & = &  < T^{0 0}>  \nonumber \\
                & =  & \frac{1}{\pi W} \sum_n \Omega(k^2_n)
                                                              \Big[ 2 \Sigma_v^0 - \Omega (k_{n}^2-a^2) - a^2  S(k_n) \Big]
                                                              \nonumber \\
                &       &   \qquad +  \frac{1}{2 \, G}(\Sigma_{v}^{0})^{2}    \, ,
\end{eqnarray}
where $\Omega(x) = \sqrt{\Lambda^2-x}$, and
\begin{eqnarray}
 P & = & < T^{z z}> \nonumber \\
    &  = & \frac{-1}{\pi W}\sum_n \left[ \Omega (k_{n}^2)
                                                          \Omega (k_{n}^2-a^2)
                                                           -a^2 S(k_n)
                                                           \right] \nonumber \\
     &  & \qquad - \frac{1}{2 \, G} \, (\Sigma_{v}^{0})^{2} \, \Sigma_s^2
\end{eqnarray}
where $z$ is the longitudinal direction of the GNR.
Then, the expression for the chemical potential reads
\begin{equation}
\mu = \frac{G \, \Lambda^2}{\pi^2}f(x) \, -\, \sqrt{\Lambda^2+\Sigma_s^2} = -
W \!\! F,
\label{wf}
\end{equation}
where $W \!\! F$ is the work function.
As a side remark, according to reference \cite{march}, the chemical potential
computed in the way described above, contains already both the bulk and surface
contributions from the energy density and pressure.

Graphene is recovered in the limit $W \rightarrow \infty$ and, in this case,
$\,x=\Lambda W/\pi$ also diverges, while $f(x \rightarrow \infty) = \pi/2$.
Therefore, for graphene if one sets $G = G_{crit}$ and $\Sigma_s = 0$,
one defines the parameters of the model.
Then, using $G_{crit}$, it follows from Eq. (\ref{wf}) that $\Lambda = 2W \!\! F$. Given that
the experimental value for the graphene work function is $4.8$ eV \cite{suzuki0,suzuki},
with determine that $\Lambda = 9.6$ eV.
From now on, we will use always these results for computing the nanoribbon gaps within the
effective relativistic field theoretical model given by the lagrangian density (\ref{lagrangiana}).

The above definitions fully parameterizes Eqs. (\ref
{sigmas})-(\ref{sigmavtheta}), allowing the computation of a
self-consistent solution for $\Sigma_s$ as a function of the
nanoribbon width $W$. Furthermore, given a value for $\Lambda$, one
can compute the surface density $\sigma= \Lambda^{2} / \pi$ .

\section{Results}
In order to solve the model numerically, besides $G_{crit}$ and $\Lambda = 9.6$ eV, we use
$v_F=c/300$. \cite{castroneto} Our numerical procedure is as follows:
{\it i)} For a given $W$, $x=\Lambda W/\pi$ and $k_{n}=n\pi/W$ are defined, constraining
the integer $n_{max} \le x $; {\it ii)} given that $\Sigma^\theta_v = 0$,
$a^2 = k^2_n + \Sigma^2_s$  and Eq. (\ref{sigmas}) can be solved  iteratively to compute $E_{g}= 2 \Sigma_{s}$.
$\Sigma_{v}^{0}$ is calculated directly from Eq. (\ref{sigmav0}).

In Fig. \ref{fig1} the results of the model for the band gap $E_g$
as a function of the nanoribbon width are compared with the
experimental data of reference \cite{xiaolin}. As observed in the
figure, the model reproduces the GNRs experimental gap for a large
range of nanoribbon widths. We call the reader attention, that the
curve in Fig. \ref{fig1} is the direct outcome from the model
without any \textit{ad hoc} normalization to reproduce the right
$E_g$ scale. This striking agreement between experimental data and
theoretical predictions seems to indicate that the model, as defined
above, captures the essential physics required to understand the gap
formation in nanoribbons. Moreover, the results summarized in Fig.
\ref{fig1} give support to the idea that effective interacting
valence Dirac electrons which generates dynamically a band gap is
useful to describe the electronic properties of GNRs.

The experiments reported in Ref.\cite{suzuki0} and \cite{suzuki} show that for CNTs with radius between 0.5 and 1.5 nm,
the work function oscillates with an amplitude of $\sim 0.5$ eV or less around the graphene work function
$W \!\! F$.
As the CNT radius increases, the oscillations rapidly become very smooth.
This consideration may help in understanding why our model still works for GNRs with small widths.
Moreover, it also shows that the graphene scale, used in the definition of our model, is robust regarding the
GNR finite-size details. Possibly, the massless fermions picture reinforces this point, since the localization of
carriers is strongly suppressed favoring ballistic behavior for them.

\begin{figure}[t]
\vspace{-0.3cm}
\includegraphics[scale=0.3,angle=0]{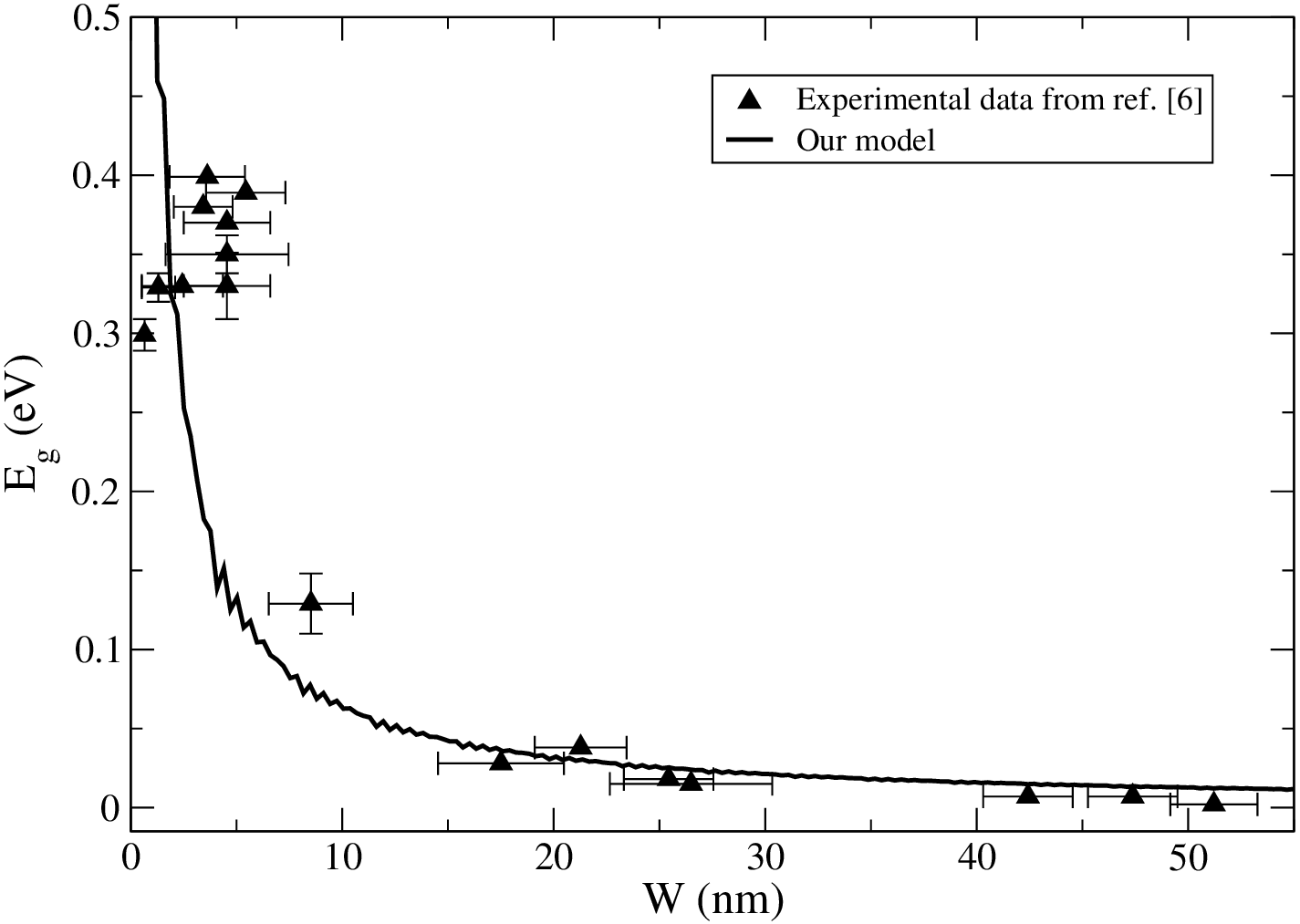}
\vspace{-0.3cm} \caption{ GNR bandgap as a function of ribbon width
$W$.} \label{fig1}
\end{figure}

Let us now compare the model with other theoretical predictions. In Ref. \cite{ywson},
first-principles calculations for three types of armchair-edged GNRs have been reported.
Their results are similar to ours. For smaller widths, previous theoretical calculations\cite{dft-vbarone} have been done using DFT functionals from the works of Perdew et al. (PBE)\cite{Perdew} and Heyd et al. (HSE)\cite{Heyd}.  According to the authors, the bandgap oscillates as a function of
$W$. A similar behavior is observed in our model -- see Fig. \ref{fig2}. Typically, our results are in between those of the two DFT (PBE, HSE) calculations.

From the point of view of the relativistic model, the oscillations
in the physical quantities are related to $n_{max}$. As the width W
decreases,  $x=\,\Lambda W/\pi$ decreases and higher values for the
integer $n$ become possible. Whenever $n_{max}$ increases by unity, due to the behavior of the derivative of $N$ relative to the
single particle energy, a Van-Hove singularity shows up. This
explains the oscillatory behavior of the GNR bandgaps observed in
our model. As seen in Fig. \ref{fig2}, whenever $W$ gets larger, the
amplitude of oscillations diminishes and they become sizable only
for a width of the order of a few nanometers or smaller.

\begin{figure}[t]
\includegraphics[scale=0.3,angle=0]{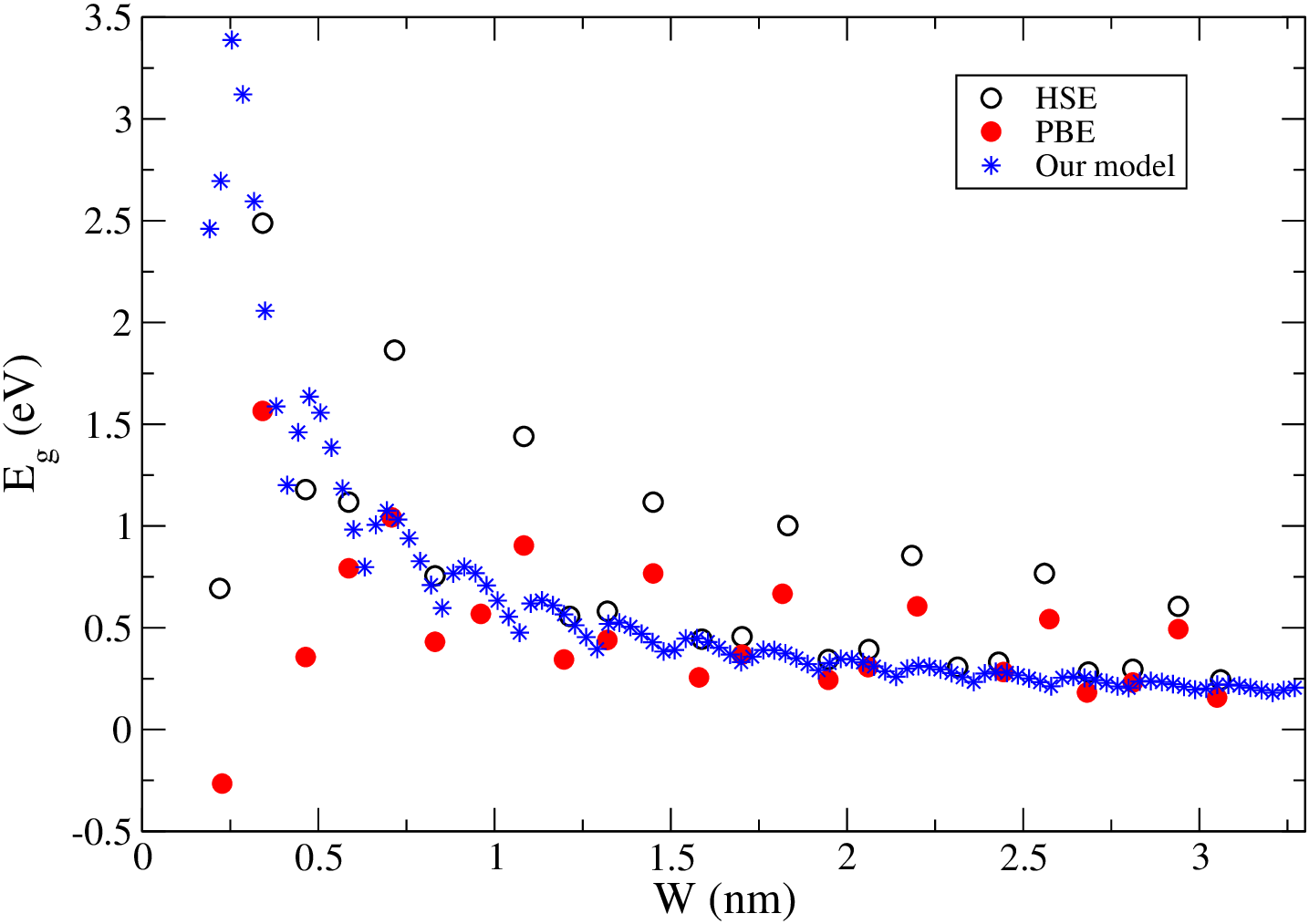}
\vspace{-0.3cm} \caption{ GNR bandgap as a function of ribbon width.
Theoretical calculations from PBE and HSE density functionals
\cite{dft-vbarone}. } \label{fig2}
\end{figure}

\section{Concluding Remarks}

Our results for the effective relativistic field theoretical model show that it accounts reasonably well for some of the physics
require to understand the GNR from small to large widths.
Moreover, the graphene scale is used to defined the model parameters and it seems to be a robust scale
regarding the GNR finite-size details.
In this sense, our results are very auspicious for further investigations of GNRs with the present model
which may, as claimed before, represent a first order approximation to a DFT calculation.
A nice feature of the relativistic field approach is its simplicity. Indeed, it allows for a detailed analytical
description that, eventually, can guide our intuition in the investigation of GNRs and in looking for general driving
physics trends.

To conclude, we would like to make general considerations on how to improve the present effective model.
For small widths, due to the increase on the kinetic energy, some of the fermions can jump into the conduction band.
Therefore, for small $W$'s the model instead of having a single cut-off $\Lambda$, will require two, one for the valence band $\Lambda_1$ and another for the conduction band $\Lambda_2$.
The presence of the new scale could lead to larger amplitude oscillations as the \textit{ab initio} calculations
suggests -- see figure \ref{fig2}.
Moreover, when $W$ becomes smaller, the edges distinguish armchair and zigzag GNRs, as discussed in Ref. \cite{fertig}.
This requires refined boundary conditions for the spinors at the GNR at the boundaries.
In general $\Sigma_v^{\theta}$ vanishes due to the symmetry between $k_n$ and $- k_n$. However, if this symmetry is
broken due to the boundary conditions, then one should solve Eqs.(\ref{sigmas})-(\ref{sigmavtheta})
self-consistently. In any case, the relativistic model still keeps its simplicity and the way we have solved the model
is, essentially, unchanged.

\section*{Acknowlegments}

The authors express their thanks to Conselho Nacional
de Desenvolvimento Cient\'{\i}fico e   Tecnol\'ogico  (CNPq),
Coordena\c c\~ao de Aperfei\c coamento de N\' ivel Superior (CAPES) and to
Funda\c c\~ao de Amparo \`a Pesquisa do Estado de S\~ao Paulo
(FAPESP) for partial financial support.

\end{document}